\documentclass[conference,compsoc]{IEEEtran}
\IEEEoverridecommandlockouts

\usepackage{graphicx}
\usepackage{adjustbox}
\usepackage{float}
\usepackage{enumerate}
\usepackage{multicol,multirow, tabularx}
\usepackage[table]{xcolor}
\usepackage{amsmath}
\usepackage{amsthm}
\usepackage{amsfonts}
\usepackage[para]{threeparttable}
\usepackage{stfloats}
\usepackage{subcaption}
\theoremstyle{definition}

\newtheorem{definition}{Definition}

\theoremstyle{definition}

\usepackage{algorithm, algpseudocode}

\algnewcommand\algorithmicforeach{\textbf{for each}}
\algdef{S}[FOR]{ForEach}[1]{\algorithmicforeach\ #1\ \algorithmicdo}

\ifCLASSOPTIONcompsoc
  \usepackage[nocompress]{cite}
\else
  \usepackage{cite}
\fi
\ifCLASSINFOpdf
\else
\fi

\begin{document}

\title{CERMET: Coding for Energy Reduction with Multiple Encryption Techniques\\ \textit{It's easy being green}
\thanks{J.~Woo, V.~Adat Vasudevan and M.~Médard are with Massachusetts Institute of Technology, USA (email: \{jc\_woo, vipindev, medard\}@mit.edu). B.~Kim is with University of Illinois Urbana-Champaign, USA (e-mail: bdkim4@illinois.edu). A.~Cohen is with Technion, Israel (e-mail: alecohen@technion.ac.il). R.~G. L. D'Oliveira is with Clemson University, USA (e-mail: rdolive@clemson.edu). T.~Stahlbuhk is with MIT Lincoln Laboratory, USA (e-mail: thomas.stahlbuhk@ll.mit.edu).}
}

\author{
    \IEEEauthorblockN{Jongchan Woo, Vipindev Adat Vasudevan, Benjamin Kim, Alejandro Cohen, \\Rafael G. L. D'Oliveira, Thomas Stahlbuhk, and Muriel M\'edard}
}
\maketitle

\begin{abstract}
This paper presents CERMET, an energy-efficient hardware architecture designed for hardware-constrained cryptosystems. CERMET employs a base cryptosystem in conjunction with network coding to provide both information-theoretic and computational security while reducing energy consumption per bit. This paper introduces the hardware architecture for the system and explores various optimizations to enhance its performance. The universality of the approach is demonstrated by designing the architecture to accommodate both asymmetric and symmetric cryptosystems. The analysis reveals that the benefits of this proposed approach are multifold, reducing energy per bit and area without compromising security or throughput. The optimized hardware architectures can achieve below 1 pJ/bit operations for AES-256. Furthermore, for a public key cryptosystem based on Elliptic Curve Cryptography (ECC), a remarkable 14.6× reduction in energy per bit and a 9.3× reduction in area are observed, bringing it to less than 1 nJ/bit.
\end{abstract}


% \fi
%\IEEEpeerreviewmaketitle

\section{Introduction}
The cost of encryption and decryption depends on multiple factors, such as the cryptographic algorithm used, the specific hardware and software implementations, and the amount of data being encrypted. In most cases, the implementation of a strong cryptosystem also requires complex mathematical operations in designated finite fields. The possibility of an energy-efficient hardware implementation is an important consideration when selecting a cryptosystem for practical purposes. Many standard cryptosystems such as the Advanced Encryption Standard (AES) \cite{daemen1999aes} have been extensively studied for efficient hardware implementations \cite{8721457, 1690090,Mathew201053GbpsNG,7019004,7573553, 10.1007/978-3-540-74735-2_16,  10.1109/TIFS.2018.2869344} but novel cryptosystems such as elliptic curve cryptography (ECC) \cite{Hankerson2004GuideTE} and many post-quantum security schemes still require a large amount of energy per bit for efficient implementations \cite{article} making them unfeasible for resource-constrained devices.

With the widespread growth of the Internet of Things (IoT), the security of resource-constrained devices has been of profound interest. This requirement will only continue with the emergence of 5G massive Machine Type Communication (mMTC) \cite{IMT2020vision}, which addresses the unique requirements of large-scale IoT networks. However, a significant challenge arises from the fact that most IoT devices, such as sensors, are resource-constrained, possessing limited memory and computational capabilities. Despite these limitations, these devices often handle sensitive data, including personal information, location data, and sensory information from critical infrastructures \cite{makhdoom2018anatomy}. Ensuring the confidentiality and security of such data is of utmost importance, yet it proves to be extremely challenging given the constraints of these devices \cite{williams2022survey,adat2018security}. Traditional encryption schemes are often unfeasible due to the low energy budget of many IoT devices. Consequently, a significant amount of research has been dedicated to studying efficient lightweight cryptographic protocols suitable for resource-constrained devices \cite{Thakor2021,tewari2017cryptanalysis,mckay2017report}. 

On the other hand, from a broader system view, the energy requirement of cryptosystems was not the dominant factor in a reliable communication system until recently. To ensure reliable and secure communication over unreliable channels, modern communication systems depend on channel coding and cryptography. There have been constant efforts to propose and standardize coding and cryptographic schemes to be used for communication since the beginning of information theory. There exists a large number of such schemes in both regimes, and each one of them requires a specific set of operations, which also requires specific hardware/circuit design, at both the transmitter and receiver sides. In the past, decoding for error correction required significantly higher energy per bit requirement compared to standard decryption schemes such as the Advanced Encryption Standard (AES) \cite{daemen1999aes}. However, there have been recent works showcasing a universal decoder, known as Guessing Random Additive Noise Decoding (GRAND) \cite{duffy2019capacity,duffy2022ordered}, which can decode any coding scheme by the work of guessing the noise patterns with very low energy consumption. \cite{riaz2022universal} presented a chip that could reduce the energy per bit for decoding to less than 1 pJ/b, making the decryption (considering a standard cryptosystem such as AES) the dominant factor in the energy analysis of the receiver. As AES-128 is increasingly perceived as vulnerable, the National Institute of Standards and Technology (NIST) has initiated the standardization process for post-quantum cryptographic protocols. In the interim, NIST recommends AES-256 as a more secure alternative \cite{NIST-PQC}. The state-of-the-art AES decryption implementations \cite{1690090, Mathew201053GbpsNG, 7019004, 7573553} exhibit significantly higher energy requirements, amplifying the emphasis on energy-efficient cryptosystems in ongoing research.

%Substituting a stronger cryptographic scheme with a weaker one is not a favorable solution for reducing energy consumption. Another factor that affects energy consumption is the amount of data being encrypted per block.

The possibility of a universal decryption, such as GRAND for decoding, is not very promising, since each cryptosystem has its own keys and structure of operations.  However, a hybrid and universal cryptosystem using network coding has been proposed in \cite{cohen2021network}. This approach, called Hybrid Universal Network Coding Cryptosystem (HUNCC), uses any existing cryptographic protocol as a component and reduces the computational complexity of the cryptographic operations. This is achieved by applying coding schemes before encryption. HUNCC efficiently combines information-theoretic approaches, such as network coding, with cryptographic schemes to provide individual secrecy for messages. HUNCC ensures individual secrecy for the messages by encoding them using a linear coding scheme and encrypting a fraction of the outgoing messages. Unless an eavesdropper is able to decrypt the encrypted part of the data, it will not learn anything about any individual message. In essence, this scheme achieves a high communication rate without compromising the security of the cryptosystem. 

% Since linear coding operations are computationally less expensive, this new approach can also be a viable alternative for designing energy-efficient hardware implementation of strong cryptosystems within the low energy budget of IoT devices. In this work, we explore the hardware architecture of an energy efficient cryptosystem based on HUNCC and discuss the necessary implementation details. Furthermore, a detailed comparison of the proposed architecture, called CERMET (Coding for Energy Reduction with Multiple Encryption Techniques), with traditional security schemes is also presented. Using HUNCC, we show that that the AES-256 decryption can be realized within a 1pJ/bit energy budget, making it suitable for the IoT devices and bringing it to the same scale as the decoding operations in the system. Furthermore, the proposed CERMET framework can be extended to more energy-intensive cryptosystems where we achieve a significant 9.3× reduction in the energy per bit for decryption for a public key cryptosystem based on ECC. The main contributions of this paper are the following.

Since linear coding operations are computationally less expensive, this new approach can also be a viable alternative for designing energy-efficient hardware implementation of strong cryptosystems within the low energy budget of IoT devices. In this work, we explore the hardware architecture of an energy-efficient cryptosystem based on HUNCC and discuss the necessary implementation details. Furthermore, a detailed comparison of the proposed architecture, called CERMET (Coding for Energy Reduction with Multiple Encryption Techniques), with traditional security schemes is also presented. In the case of AES-256, CERMET achieves a decryption energy efficiency of less than 1pJ/bit, representing a 2.5× improvement compared to the state-of-the-art AES-256 implementation \cite{8721457}  under identical simulation conditions. This achievement aligns the energy requirements of AES-256 decryption with those of decoding operations in the system and makes it suitable for IoT devices. Furthermore, when extended to more energy-intensive cryptosystems, such as a public key cryptosystem based on ECC, CERMET achieves a significant 9.3× reduction in the energy per bit for decryption. The main contributions of this paper are the following.

\begin{itemize}
    \item The paper introduces an energy-efficient hardware implementation of HUNCC with AES as the underlying cryptographic protocol. The energy analysis of the proposed cryptosystem is conducted with hardware synthesis and compared to traditional AES approaches.
    
    \item Trade-offs between energy per bit, area, throughput, security, and number of channels are explored, along with different optimization techniques in hardware design.
    %\item We discuss the different assumptions from the information-theoretic aspect of the cryptosystem and analyze the impact of input uniformity on security. We also use the novel mutual information estimator using neural networks (MINE) to check the information leakage because of the partial encryption in HUNCC and present the necessary guidelines for its practical use.
    %\item The universality of HUNCC by implementing it with a symmetric cryptosystem (AES 256) and an asymmetric cryptosystem (ECC). We also show the significant improvement that HUNCC can provide in terms of energy per bit for decryption in an otherwise computationally expensive cryptographic protocol such as a public key cryptosystem. 
    \item The universality of HUNCC is demonstrated by implementing it with both a symmetric cryptosystem (AES 256) and an asymmetric cryptosystem (ECC). The significant improvement that HUNCC can provide in terms of energy per bit for decryption in an otherwise computationally expensive cryptographic protocol such as a public-key cryptosystem is highlighted. 

    \item The modularity of the CERMET framework and its ability to adapt to different cryptosystems are explained in detail. The flexibility in adopting different hardware techniques allows the proposed framework to cater to specific implementation requirements and provide a scalable solution for a wide range of applications.
\end{itemize}

The rest of the paper is organized as follows. Section \ref{sec:SOTA} offers an overview of our security definition and the energy requirements of cryptosystems. Section \ref{sec:system} presents the proposed architecture and essential modules, followed by different hardware design approaches in Section \ref{hardware_design}. The performance evaluation and results from our implementations are presented in Section \ref{section:evaluation and results}. Section \ref{discussions} provides additional comments and inferences about the proposed architecture followed by conclusions in Section \ref{conclusions}. 

\section{Previous Work} \label{sec:SOTA}

\subsection{Security Definitions} \label{SOTA:Security}

In 1948, Claude Shannon published the seminal work \cite{shannon1949communication} which introduced the concept of perfect security for cryptographic schemes.\footnote{This publication built upon an earlier, classified version he had written in 1945 \cite{shannon1945mathematical}. Interestingly, this work precedes another of Shannon's influential papers \cite{shannon1948mathematical}.} A cryptographic scheme achieves perfect security when the mutual information $I(M;X)$ between the message $M$ and the received signal $X$ is equal to zero, meaning that $M$ and $X$ are statistically independent. Considering the widely recognized Alice-Bob-Eve model, perfect security can only be obtained if both Alice and Bob share a random key with an entropy as large as the message.

The requirement for Alice and Bob to share large secret keys is often impractical. As a result, significant efforts have been directed towards finding alternatives that relax on the perfect secrecy condition. One such method assumes that potential eavesdroppers have limited computational capabilities. Schemes relying on this assumption are known as \emph{computationally secure} and rely on the existence of one-way functions which are hard to invert. In this setup, Alice uses a one-way function to encrypt her private message before sending it to Bob. Bob is then able to efficiently invert the message, with the help of a secret key, while an eavesdropper without access to the secret key is not able to do so due to her computational constraints.

\begin{definition}[Symmetric and Asymmetric Cryptosystems]\label{Crypto_scheme}
A symmetric/asymmetric cryptosystem consists of:
\begin{enumerate}
    \item A key generation algorithm $\mathrm{Gen}(\kappa)$ which takes as input a security parameter $\kappa$ and generates a public key $p_k$ and a secret key $s_k$. In the case of the symmetric cryptosystem, the public and secret keys are the same, i.e., $p_k=s_k$.

    \item An encryption algorithm $\mathrm{Enc}(m,p_k)$ that takes as input a message $m$ belonging to some set of messages $\mathcal{M}$ and the public key $p_k$ and then outputs a ciphertext $c$ belonging to some set of ciphertexts $\mathcal{C}$.

    \item A polynomial-time decryption algorithm $\mathrm{Dec}(c,s_k)$ which takes as input a ciphertext $c=\mathrm{Enc}(m,p_k)$ and the secret key $s_k$ and outputs the original message $m$.
\end{enumerate}
\end{definition}

The key assumption of a computationally secure cryptosystem is that decrypting the encrypted message without the secret key is computationally infeasible. The literature has proposed various ways to mathematically characterize this impracticality. One popular way is through the notion of security level \cite{lenstra04}. A cryptosystem has security level $b$, termed as being $b$ bit secure, if the amount of operations expected to decode the encrypted message, without knowledge of the secret key, is~$2^b$. There are also more theoretical definitions like that of ciphertext indistinguishability \cite{katz2020introduction}, where an attacker cannot determine any information about the plaintext by observing the ciphertext, even when given some auxiliary information.

Alongside advances in computational security, another deviation from perfect privacy has been extensively explored in academic circles, mainly by information theorists \cite{wyner1975wire,ozarow1985wire,el2012secure,liang2009information,bloch2011physical}. Instead of restricting the computational power of the eavesdropper, in \emph{physical layer security} one limits how much information Eve can obtain about the encrypted message.

In this paradigm, Wyner \cite{wyner1975wire} introduced the notion of the wiretap channel, the physical layer analogue of the Alice, Bob, and Eve trio, and showed that perfect security can be obtained, though at a lower communication rate. For example, in \cite{ozarow1985wire} it is assumed that Eve can observe any subset of size $w$ from a total of $n$ transmitted symbols. Denoting this set by $Y_{E_w}$, it was shown that there exist codes with communication rate $\frac{n-w}{n}$ that are perfectly secure, i.e., $I(M;Y_{E_w}) = 0$. To achieve perfect secrecy in physical layer security, one must pay the price of decreasing the communication rate.

To improve efficiency with respect to the communication rate, a further deviation from Shannon's perfect secrecy was introduced: the concept of \emph{individual secrecy} \cite{kobayashi2013secure,bhattad2005weakly,silva2009universal,mansour2014secrecy,chen2015individual,goldenbaum2015multiple,chensecure,cohen2018secure}. This concept is more intuitively understood within a network scenario. Consider Alice wanting to transmit multiple messages to Bob, labeled $M_1, \ldots, M_m$, through $m$ transmissions $X_1$, \ldots, $X_m$ of which Eve observes at most $w$. The notion of individual secrecy guarantees that $H(M_i|Y_{E,w}) = H(M_i)$, for all $i = 1, \ldots, m$. This is generally weaker than perfect secrecy, where $H(M_1,\ldots, M_m |Y_{E,w}) = H(M_1,\ldots, M_m)$. Simply put, while Eve obtains some information about the correlations between the messages, she obtains no information about any individual message. In turn, this allows for a significant boost in the communication rate, making encryption cost-free in terms of communication rate. In \cite{cohen2021network, cohen2022partial}, individual secrecy analogs for computational security were proposed for both security level and ciphertext indistinguishability.

\begin{figure}[!t]
    \centering
    \includegraphics[width=1\linewidth]{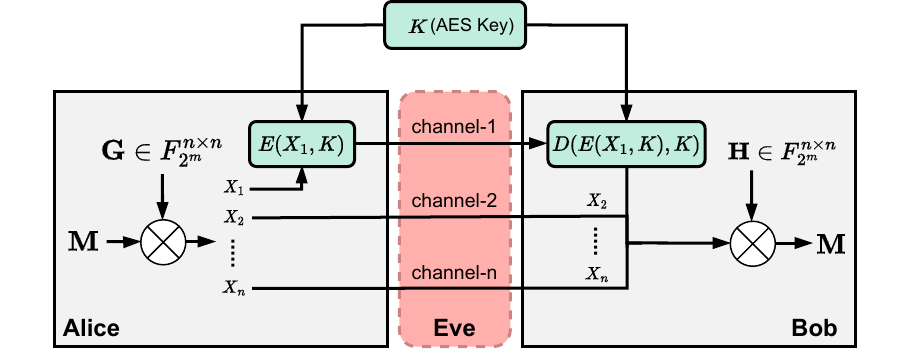}
    \caption{High level overview of HUNCC}
    \label{fig:HUNCC_overview}
\end{figure}

\subsection{Energy requirement of Cryptosystems} \label{SOTA:Energy}
%Discussion on different cryptosystems – symmetric and asymmetric – AES – ECC – Post-quantum cryptography - Why Energy/bit, power, area and throughput are our matrices

As discussed in Subsection \ref{SOTA:Security}, there are a wide variety of cryptographic protocols, each with its own advantages and disadvantages. One factor that impacts the selection of a particular protocol for an application is the cost vs. security tradeoff. Asymmetric key cryptosystems provide higher-level security at the expense of more complex operations compared to symmetric key cryptosystems \cite{stinson2005cryptography}. This is one of the major reasons why public key cryptosystems are traditionally used to share a secret key for further long-term communication. Both the implementation cost and the operational cost are significant concerns when selecting a cryptosystem. The Rijndael algorithm, used in AES \cite{daemen1999aes}, became popularized due to its efficient implementation in both hardware and software, along with its strong security and flexible key size.

The hardware implementation cost corresponds to the complexity of operations and the number of logical operations required. The chip's area, which is significant for manufacturing cost, is proportional to the number of logic gates required to implement the module, making it a commonly considered metric for evaluation. Multiple optimization techniques, such as serialization of operations and repeated use of identical modules, have been employed to optimize the area. However, this may result in lower throughput, an important parameter for operational efficiency. In addition, power consumption and energy per bit are two major indicators of operational cost. The power consumption can be reduced by lowering throughput, but this can reduce the energy efficiency of the system. The energy per bit is a parameter that is commonly used to compare hardware performance since it takes both power consumption and throughput into account. In this work, we discuss all these parameters and their trade-offs while evaluating our system. We also propose optimizations and suitable alterations to a base model to achieve the best trade-off between different key performance indicators.

Significant research has been dedicated to optimizing the performance of standard cryptosystems in hardware. In the case of symmetric cryptosystems such as AES, various hardware technologies have been leveraged to achieve optimal implementations over the years \cite{8721457, 1690090,Mathew201053GbpsNG,7019004,7573553}. Notably, one of the most efficient AES implementations achieved an impressive energy efficiency of 4.08 pJ/bit on 65 nm CMOS process fabricated chip \cite{8721457}. Efforts have also been made to implement energy-efficient asymmetric key cryptosystems with higher security. However, even the most efficient public key cryptosystems, using Elliptic Curve Cryptography (ECC), still exhibit energy per bit performance in the range of $\mu$J/bit \cite{8662528}. This is significantly higher compared to AES. Moreover, ECC cryptosystems require considerably more time for decryption operations, resulting in lower throughput. In this work, we specifically focus on these two cryptosystems, highlighting the universality of HUNCC over both symmetric and asymmetric cryptosystems. We aim to showcase how HUNCC can improve the above-mentioned Key Performance Indicators (KPIs) for both AES and ECC, offering a promising approach for enhanced energy-efficient cryptographic implementations.

%\subsection{HUNCC}\label{subsec:huncc} %Alejandro

The recently proposed HUNCC scheme \cite{cohen2021network,d2021post} could provide an interesting perspective to the study of energy efficient cryptosystems. This system combines information-theoretic security with a computationally secure cryptosystem to achieve a computationally efficient secure scheme at high data rates (see \cite[Theorem 3]{cohen2021network}). As illustrated in Figure 1, the secure network-coding scheme operates by first linearly premixing $n$ messages using a generator matrix $\textbf{G}\in F^{n \times n}_{2^m}$ of a specific type of network coding scheme \cite{silva2009universal,cohen2018secure}. Subsequently, the mixed plaintext is partially encrypted using any cryptosystem. These partially encrypted, premixed messages are then transmitted over multiple channels in an untrusted network, such as a multipath network with $n$ paths. Surprisingly, the hybrid scheme maintains the same computational security level $b$ as suggested by the cryptosystem (see \cite[Theorem 1]{cohen2021network}), even if only a single channel (out of the $n$ channels in the network) is encrypted. As the linear coding operations are less complex compared to the cryptographic operations, the proposed cryptosystem can achieve this with significantly lower energy consumption. Furthermore, it provides information-theoretic security against an eavesdropper who does not observe all communication channels. HUNCC can be applied to any communication network and to any cryptosystem making it suitable for resource-constrained devices. However, detailed analysis and implementation of such a hybrid system on hardware is little explored. Furthermore, a quantitative analysis and energy profiling of the system is required for its comparison with the state-of-the-art cryptographic solutions. This work addresses this gap by providing a detailed architectural framework for the coded cryptosystem and different techniques for optimized performance.

\section{System Architecture} \label{sec:system}
%Receiver side of HUNCC – Discussion on the structure (crypto core and unmixing process) – Consider AES 128 case – Unmixing requires less energy compared to decryption up to a number of channels

In this section, we discuss the architectural design and components of the coded cryptosystem. Although this approach can be used with any cryptosystem, we consider AES-256 as the cryptographic algorithm in our setting and Random Linear Network Coding (RLNC) as the coding scheme for the architecture description. All the operations involved in the sender and receiver sides are symmetric. However, we focus on the receiver side for the hardware architecture design, since we also analyze the throughput of the system and include the intermittent output registers while calculating the energy and area.

%@Alejandro: May be you can rewrite this paragraph with more details of the coding scheme.\\ 
%@Vipindev: Sure. Is the code implemented that of Silva's paper that I shared? If yes. Please ask Jongchan for all the parameters of the code selected, and I will rewrite this section with the precise implementation. 
%@alejandro: Jongchan has followed the MRD code as in Silva's paper. I think he considers m = 16 in $F_q, q =2^{16}$, n=k case, $n \in {[2:15]}$, the input and output are of the same number. We can clarify this in today's meeting. 

On the sender side, the inputs to all the different $n$ channels are multiplied by a generation matrix, and one of the outgoing messages is encrypted using the cryptographic algorithm. The generation matrix $\textbf{G}$ in this work is constructed as given in \cite[Proposition 9]{silva2009universal}. For a parity check matrix $\textbf{H}\in F^{n \times n}_{2^m}$, $\textbf{G}=\textbf{H}^{-1}$. Thus, for $m\geq n$ and $h_1\ldots h_n \in F_{2^m}$ linearly independent elements over $F_{2}$, every row of $\textbf{H}\in F^{n \times n}_{2^m}$  defines a Maximum Rank Distance (MRD) code\cite{gabidulin1985theory,roth1991maximum}. In particular, in the setting considered herein for $m=16$, $\textbf{H} = [H_{ij}] \in F^{n \times n}_{2^{16}}$ is given by $H_{ij}=h_{i}^{2^{j-1}}$, for $1\leq i \leq n$ and $1\leq j \leq n$. A detailed explanation of the HUNCC algorithm from the literature is added in the appendix \ref{appendix:HUNCC} for further reference.

On the receiver side, the inverse matrix of the generation matrix $\textbf{G}^{-1}$ (i.e., the parity matrix $\textbf{H}$) is stored to decode the messages. When the receiver obtains the incoming messages, the unencrypted messages are directly fed to the registers, while the encrypted message is decrypted first using the decryption algorithm. Once the decryption is complete, the messages are decoded using the inverse matrix. Multiple levels of parallelization are possible in the process, which will be explained in the hardware design section.

%\subsection{System Architecture Overview} \label{system:overview}
\begin{figure*}%[!t]
    \centering
    \includegraphics[width=1\linewidth]{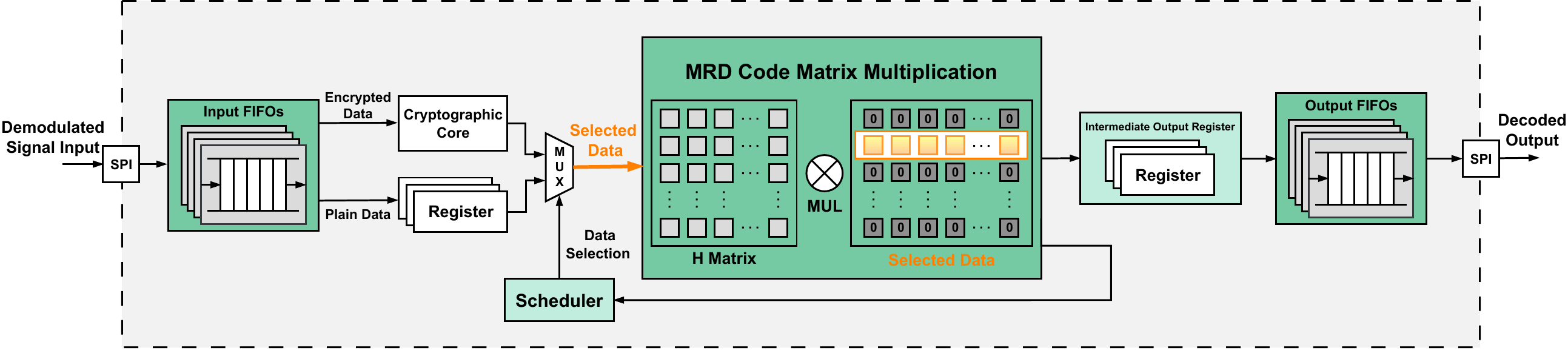}
    \caption{System architecture of the receiver}
    \label{fig:system_architecture}
\end{figure*}

Figure~\ref{fig:system_architecture} provides a comprehensive view of the system architecture on the receiver side, which includes several stages for handling and processing data. The architecture is designed to handle different channel inputs, decrypt the encrypted portion of the input, and unmix decrypted and unencrypted input using the MRD code matrix, which completes the process. All these stages operate in parallel and are pipelined to maximize throughput. In this work, we focus on investigating a hardware architecture for the receiver side in HUNCC applications that is both energy and area efficient.

\subsection{Channel Inputs} \label{system:channel}

The architecture manages different channel inputs based on their encryption status. Each data path in this context corresponds to a communication channel as seen in practical multi-channel communication systems. Encrypted data is directed towards the cryptographic accelerator for decryption, while plaintext data is directly stored in registers, ensuring efficient data handling. The proposed system can have any number of input channels only bounded by the field size of the coding operations such that the maximum number of channels $n \leq m$ when the field size is $2^m$ \cite{silva2009universal}. For practical applications, increasing the field size will result in increased computational requirements, especially an exponential increase in the case of multiplication operations \cite{angelopoulos2011energy}.

\subsection{Data Selection} \label{system:data}

The architecture processes one data unit at a time during matrix multiplication, a choice made for area and energy efficiency. To manage this, a scheduler operating on a First-In-First-Out (FIFO) basis is incorporated to buffer multiple arrivals.

The system's pipelined architecture enables dynamic data selection and preparation. As soon as a data unit is fetched from an input and begins processing, the next data unit is simultaneously selected and prepared for processing. This concurrent operation minimizes delays and maximizes the system's efficiency, ensuring that the data flow remains continuous and the processing resources are optimally utilized. This dynamic data handling is a key feature of the architecture that contributes to its high performance and energy efficiency.

\subsection{MRD Code Matrix Multiplication} \label{system:multiplication}
Data from the cryptographic accelerator (decrypted data) or data paths (plain data) is sent to the matrix multiplication block. The matrix \textbf{H}, used on the receiver side, defines a MRD code for each of its rows. The process of mixing the messages with the inverse of the \textbf{H} matrix on the transmitter side provides a level of individual security, as described in \cite[Proposition 9]{silva2009universal}. On the receiver side, the fetched data is multiplied with the matrix \textbf{H}, processing one data unit at a time.

The cornerstone of our system's good performance is the energy-efficient design of the matrix hardware, which is crucial for this process. By partially decrypting the data with the cryptographic accelerator, we significantly reduce the energy consumption compared to full decryption. This approach, while maintaining the same level of security, enables a highly energy-efficient system. The detailed design and performance characteristics of this energy-efficient matrix hardware will be elaborated in Section \ref{hardware_design}.

The dimensions of the \textbf{H}-matrix are dictated by the number of channels, $n$, resulting in an $n$ by $n$ matrix. The data matrix, fetched from either the cryptographic accelerator or the plain data register, has dimensions of $N = k_{in}/m$, where $k_{in}$ is the bit length of input  (e.g., 128/16 for 128-bit AES data and GF($2^{16}$)). In the data matrix, if we fetch the i-th data, all elements of the matrix become zeros, except for those in the i-th row.

\subsection{Intermediate Output Register and Final Output} \label{system:output}

% Second paragraph of MRD code mult. can be moved to here
The multiplication process in HUNCC operates on one data unit at a time. Given that all data (crypto and plain) must be multiplied with the $\textbf{H}$ matrix to obtain the final decoded output, it is necessary to store the intermediate results. This is achieved using intermediate registers, which hold each multiplication result. This requirement is unique to HUNCC due to the premixing of all data before partial encryption. In conventional cryptosystems, the decryption process directly yields the original data, eliminating the need for such intermediate storage. However, in HUNCC, we need to unmix the input to retrieve the original data, necessitating these intermediate registers. This introduces an additional power and area cost for our system that we account for.

After all data units have been fetched and multiplied with the matrix, decoding is performed. The decoded results are then transferred to the output FIFOs. This systematic approach ensures efficient data handling and processing, facilitating the retrieval of original data from the premixed and partially encrypted input.

% \begin{itemize}
%     \item Baseline System Architecture
%     \item HUNCC architecture using AES 128 or 256 as the core core cryptosystem
% \end{itemize}

\section{Energy-Efficient Hardware Design} \label{hardware_design}

This section discusses the system's integration with cryptographic engines, AES-256 and ECC, focusing on the hardware techniques aimed at enhancing energy and area efficiency. While many researchers have focused on the energy-efficient hardware design of different cryptographic systems\cite{10.1007/978-3-642-21040-2_10, tinyECC,10.1007/978-3-319-13066-8_10,inproceedings,8662528}, this work emphasizes energy-efficient design across all blocks in the system.

\subsection{HUNCC with AES Cryptography} \label{section:huncc_with_aes_accel}
Data received from the AES-256 cryptographic accelerator or registers is segmented into multiple units in the binary extension field GF($2^{m}$). The original data can be recovered by performing matrix multiplication with the MRD code \textbf{H} matrix. Each unit of data is multiplied with matrix elements, which also reside in the same binary extension field GF($2^{m}$). Given the extensive research and numerous studies on optimal designs for AES, the AES-256 implementation from \cite{8721457}, known for its low-latency, low-area, and low energy/bit characteristics, is adopted. %It offers a superior level of security compared to AES-128, courtesy of its extended key length, and has been used for top secrets as per Suite B cryptography recommendations by NSA \cite{barker2006suite}.

\subsubsection{Multiplication in the Galois Field} \label{sub:GFM} \hfill

Various approaches have been proposed for multiplication in the Galois field \cite{lidl_niederreiter_1996}. In this work, we conducted an in-depth investigation of different multiplication algorithms in hardware, comparing their power consumption and area. Based on these comparisons, the most suitable algorithm was selected and integrated into our system, tailored to the specific architecture and performance requirements. This approach ensures the optimal balance of efficiency and performance in our system.

% As discussed in the previous section, the multiplication of two elements in Galois field necessitates a significant number of logical operations. Performing all these logical operations in real-time requires a large number of logic gates to be implemented on the hardware, raising concerns for the power and area. The fundamental concept behind this algorithm is to execute multiplication one factor at a time, ensuring that the resulting polynomial at each step has a degree less than m. This method of Galois field multiplication, known as the Russian Peasant Algorithm (RPA) multiplication, has its complexity proportional to the field size. To circumvent this, another approach is to precompute the results and use a look-up table at runtime to find the results. This method reduces the number of operations required per multiplication but comes at the expense of high memory consumption. To mitigate this issue, the lookup table can be constructed using the power representation of the finite field elements, a method known as log/exponential method. The use of precomputed look-up tables often enhances performance in software implementations since a single look-up table can be used for multiple computations parallel and thus performs better than RPA multiplication which involves substantial computations. However, in hardware implementations, this parallelization is not possible and computation requires its own look-up table. Thus, for most practical field sizes, the RPA multiplication becomes more memory and power efficient than the log/exponential look-up table based operation.

As discussed in the previous section, the multiplication of two elements in Galois field necessitates a significant number of logical operations. Performing all these logical operations in real-time requires a large number of logic gates to be implemented in the hardware, raising concerns for power and area. The fundamental concept behind the Russian Peasant Algorithm (RPA) multiplication~\cite{gimmestad_1991}, detailed in Appendix~\ref{appendix:mult_algorithm}, is to execute multiplication one factor at a time, ensuring that the resulting polynomial at each step has a degree less than m. This method has its complexity proportional to the field size.

To circumvent the complexity of RPA, another approach is to precompute the results and use a look-up table at runtime to find the results~\cite{10.1007/3-540-47555-9_14}. This method reduces the number of operations required per multiplication but comes at the expense of high memory consumption. To mitigate this issue, the lookup table can be constructed using the power representation of the finite field elements, a method known as the log/exponential method, which is further explained in Appendix~\ref{appendix:mult_algorithm}.

The use of precomputed look-up tables often enhances performance in software implementations since a single look-up table can be used for multiple computations in parallel and thus performs better than RPA multiplication which involves substantial computations. However, in hardware implementations, this parallelization is not possible and each computation requires its own look-up table. Thus, for most practical field sizes, the RPA multiplication becomes more memory and power-efficient than the log/exponential look-up table-based operation.

\begin{figure}[!t]
    \centering
    \includegraphics[width=1\linewidth]{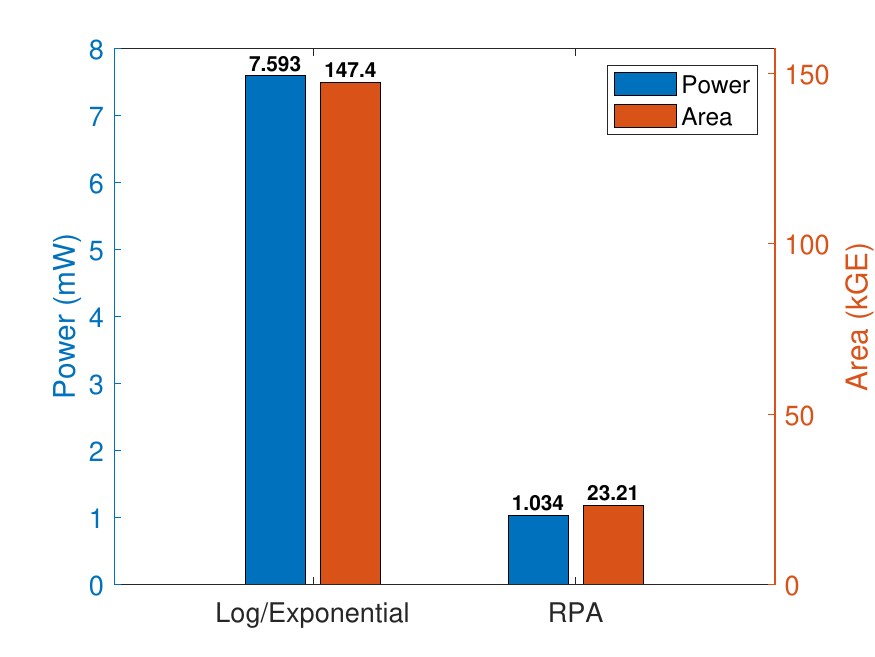}
    \caption{Power and Area comparison between two GF($2^{8}$) multiplication methods }
    \label{fig:Multiplication_Comparison}
\end{figure}

% \begin{figure}[!t]
%     \centering
%     \includegraphics[width=1\linewidth]{figures/RPA_LUT_mult_comparison.pdf}
%     \caption{Power and Area comparison between two GF multiplication methods}
%     \label{fig:RPA_LUT_mult_comp}
% \end{figure}

Figure~\ref{fig:Multiplication_Comparison} presents a comparison of the synthesized area and the simulated power of hardware-implemented matrix multiplication in GF($2^{8}$) for a four-channel scenario. In both cases, the implementation was designed to maintain the same latency for one multiplication operation between the matrix and the fetched input data. Contrary to the performance observed in software implementations \cite{10.1007/BFb0034836, 10.1145/2093139.2093141}, the results show that the RPA-based multiplication method significantly reduces both power (7.3×) and area (6.4×) compared to the look-up table method. This outcome arises because hardware implementations require multiple look-up tables stored in registers to perform several multiplication processes and maintain low latency. The area and power required for this storage in the look-up table method become dominant factors in the overall system.

For AES in GF($2^{8}$), which has a 128-bit data length divided into 16 elements, 16 parallel multiplication modules are required, and each module requires three look-up tables (two log tables for inputs and one exponential table for outputs). On the other hand, RPA-based multiplication can be significantly simplified in hardware implementations and fully leverage the parallelization capabilities of the hardware. Therefore, even though the RPA-based multiplication module still requires 16 individual multiplication modules, the absence of look-up tables leads to a much lower area and power consumption.

This analysis can be extended to GF($2^{16}$). Despite the increase in field size, the RPA-based multiplication remains more efficient. In the log/exponential method, 82\% and 89\% of power consumption and area respectively are attributed to the look-up tables. As the field size increases to $2^{16}$ from $2^{8}$, the memory required for the look-up tables increases significantly. Specifically, it doubles the size of each value (from 8 bits to 16 bits) and increases the number of values by a factor of $2^{8}$ \cite{angelopoulos2011energy, 10.1007/BFb0034836}. In contrast, the RPA method only sees around a doubling of area with similar power consumption, as per our simulations. This stark difference makes the log/exponential methods much less efficient than the RPA-based method, particularly in higher field sizes. The absence of look-up tables in the RPA-based approach continues to result in lower area and power consumption, making it a more viable option for hardware implementations that prioritize energy and area efficiency.

\subsubsection{Pipelining and Process Parallelization}\label{sub:Parallel}\hfill

\begin{figure}%[!t]
    \centering
    \includegraphics[width=1\linewidth]{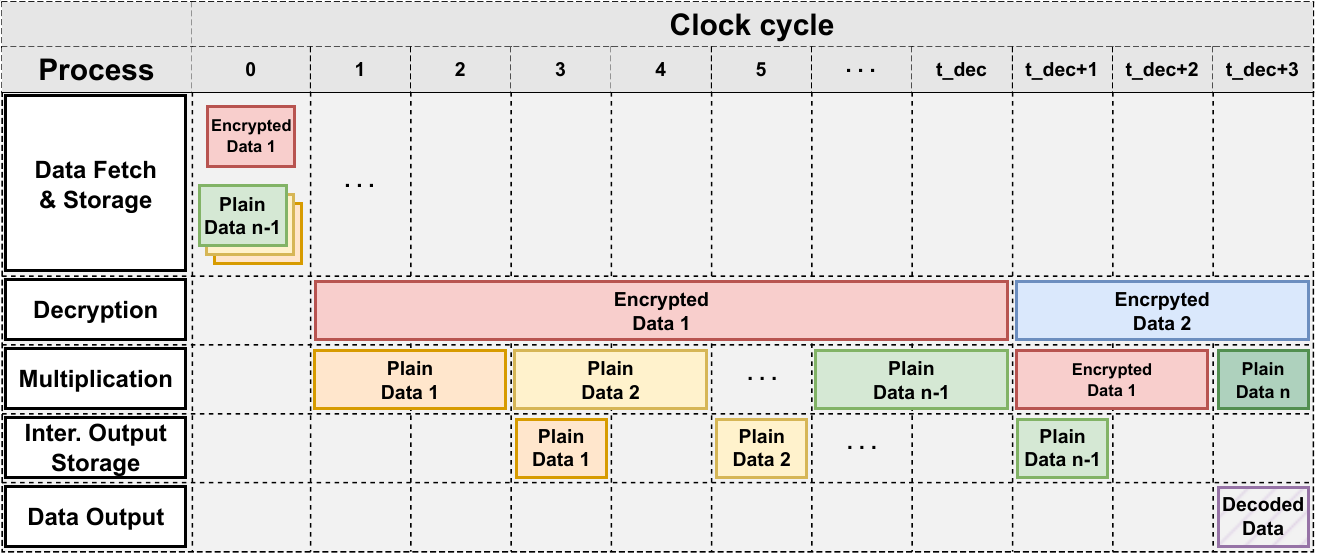}
    \caption{Pipeline diagram with  process parallelization in multiple stages}
    \label{fig:pipeline}
\end{figure}

%Beyond matrix multiplication, pipelining has been adopted throughout the entire data processing sequence, encompassing data fetch from input FIFOs, decryption in the cryptographic core, matrix multiplication, and final decoded data generation. This comprehensive integration of pipelining allows the system to operate fully pipelined, ensuring that no throughput is lost (or latency increased) compared to a baseline system. This holds true even with the additional data processing step of matrix multiplication, which is not present in the baseline system that only includes an individual cryptographic core in each channel.
Our system design incorporates both pipelining and process parallelization to enhance efficiency. There are four distinct stages in the system: Data Fetch \& Storage, Decryption, Multiplication, and Intermediate Output Storage. The final step is Data Output, which only occurs once all data have been stored in the Intermediate Output Storage.

In our implementation, the AES decryption process takes 13 (for AES-128) or 17 (for AES-256) clock cycles. During this time, plain data are being fetched and multiplication operations are running concurrently. Notably, the matrix multiplication requires 2 clock cycles, and the Intermediate Output Storage requires one additional clock cycle. This means that it takes 3 clock cycles per one data unit to be multiplied and stored in the intermediate registers. However, by pipelining these processes, we can effectively reduce this to only 2 clock cycles.

Figure~\ref{fig:pipeline} illustrates how these processes are pipelined and work in parallel. This comprehensive integration of pipelining allows the system to operate fully pipelined, ensuring that no throughput is lost compared to a conventional cryptographic system. Even with the additional data processing step of matrix multiplication, which is not present in conventional systems that only include an individual cryptographic core in each channel, our approach maintains throughput and significantly enhances the overall efficiency of our system.

 However, in scenarios with a high number of channels, where there is a large volume of data to multiply, the latency of matrix multiplication could potentially exceed that of the decryption process. This aspect will be discussed in more detail in Section~\ref{section:evaluation and results}.

\subsection{HUNCC with Energy-Intensive Cryptography} \label{sub:ECC}

The flexibility of the system allows for the integration of any cryptographic core, not just AES. This opens up the possibility of employing various cryptographic methods that were previously deemed too energy-intensive for use in encryption/decryption processes \cite{ecc_energy}. In this work, we adopt elliptic curve cryptography (ECC), a public-key cryptography approach based on the algebraic structure of elliptic curves over finite fields, for another application. Specifically, Curve25519 was chosen due to its efficient $x$-coordinate only Montgomery ladder \cite{Montgomery1987SpeedingTP}, which enables a compact architecture for the ECC hardware. In addition to adopting the energy-efficient Curve25519 hardware implementation from  \cite{9174639}, we also explored various hardware techniques to further optimize the system's performance and efficiency.

In contrast to AES, which is optimized for high-speed operation and takes under 20 clock cycles per decryption in our design, most other energy-demanding cryptographic algorithms exhibit higher latency. For instance, in our design, Curve25519 necessitates over 80k clock cycles for each elliptic curve scalar multiplication, a fundamental operation in elliptic curve-based cryptography~\cite{9174639}. This indicates that the latency bottleneck in the system is predominantly determined by the cryptographic core. To address this challenge, this study concentrates on mitigating this extended latency by leveraging a variety of architectural strategies and hardware techniques.

\subsubsection{Serialized Matrix Multiplication} \label{sub:serial}\hfill
%A) Serialized Matrix Multiplication

\begin{figure}%[!t]
    \centering
    \includegraphics[width=1\linewidth]{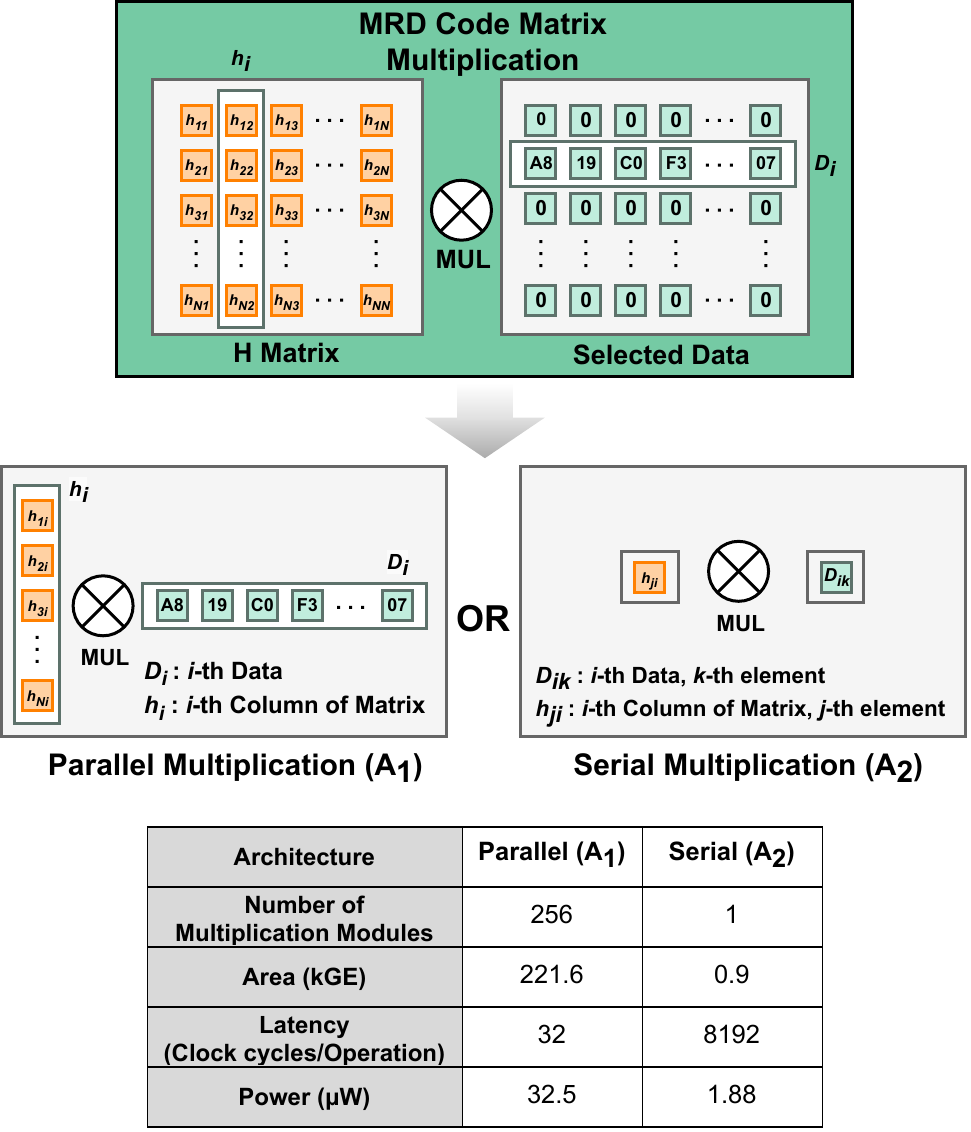}
    \caption{Comparison of multiplication architectures in 16-channel system. $A_{1}$ : serial. $A_{2}$ : parallel}
    \label{fig:matrix_multiplication_architecture}
\end{figure}

% In Section, we explored the development of hardware multiplication in the Galois field. When AES is used as the cryptographic core, it is optimized for high-speed operation, taking less than 20 clock cycles. The goal was to achieve a similar high-speed operation for matrix multiplication, leading to the creation of multiple multiplication modules operating in parallel. While this strategy effectively minimizes latency in matrix multiplication and maximizes throughput, it can become the dominant latency bottleneck in scenarios with a high number of channels.

In Section~\ref{section:huncc_with_aes_accel}, we delved into the intricacies of hardware multiplication in the binary extension field GF($2^m$). With AES as the cryptographic core, it is optimized for high-speed operation, requiring 13 clock cycles for AES-128 and 17 clock cycles for AES-256. The objective was to mirror this high-speed operation for matrix multiplication, which led to the development of multiple multiplication modules operating in parallel. Figure~\ref{fig:pipeline} illustrates how the parallel processes of decryption and matrix multiplication work. This design allows simultaneous decryption and matrix multiplication, working to minimize overall system latency and maximize throughput. 

When ECC is employed as the cryptographic core, it operates at a significantly slower speed compared to AES, thereby becoming the dominant latency factor. This discrepancy provides an opportunity to leverage the extended delay by transitioning from a parallel to a serialized structure for matrix multiplication. Rather than performing multiplication on the entire data set in one process with an array of concurrent multiplication modules, multiplication can be performed on each element (the amount of data segmented in the binary extension field) serially.

As illustrated in Figure~\ref{fig:matrix_multiplication_architecture}, once the data is fetched, it is segmented into multiple elements, each residing in GF($2^m$). For the HUNCC with AES, once the data is fetched from either plain data registers or the cryptographic core, it utilizes an array of concurrent multiplication modules, enabling the multiplication of multiple data segments in two clock cycles. However, in the case of serial multiplication, the system employs a single multiplication module, which can multiply only one data segment with one segment of the \textbf{H} matrix at every two clock cycles.

Consider a scenario with 16 channels of 256-bit long data and GF($2^{16}$). The table in Figure~\ref{fig:matrix_multiplication_architecture} compares the number of multiplication modules, area, clock cycles, and power for both parallel and serial multiplication in this scenario. For parallel multiplication, it would require 256 modules (calculated as 16, which is the 256-bit long message divided by the 16-bit field size, multiplied by the number of channels, 16). This configuration enables the system to calculate the multiplication of each data element and store the intermediate data within every 2 clock cycles, thus requiring a total of 32 clock cycles (2 times 16, which is the number of channels). However, if only one multiplication module is maintained, it would require 256 times longer than the parallelized one. Despite this, the latency is still relatively small compared to the cryptography latency.

According to the simulation with both parallel and serial multiplication cases, serial multiplication can save over 98\% of area and 94\% of power compared to the parallelized multiplication architecture. This comes at the expense of longer latency. However, the serialized multiplication can be processed within the long time of ECC core operation, making it a viable option for scenarios where power and area efficiency are of paramount importance.

In conclusion, adopting a serialized approach to the multiplication process allows for a significant reduction in power consumption within the multiplication block. Moreover, as the number of channels increases, the system area, which can become dominated by the multiplication modules, can be significantly conserved. This strategy presents a harmonious solution for power and area efficiency, underscoring the system design's inherent flexibility and adaptability.

\subsubsection{Additional Hardware Techniques} \label{sub:additional} \hfill
%B) Additional Hardware Techniques

\begin{figure}[!t]
    \centering
    \includegraphics[width=1\linewidth]{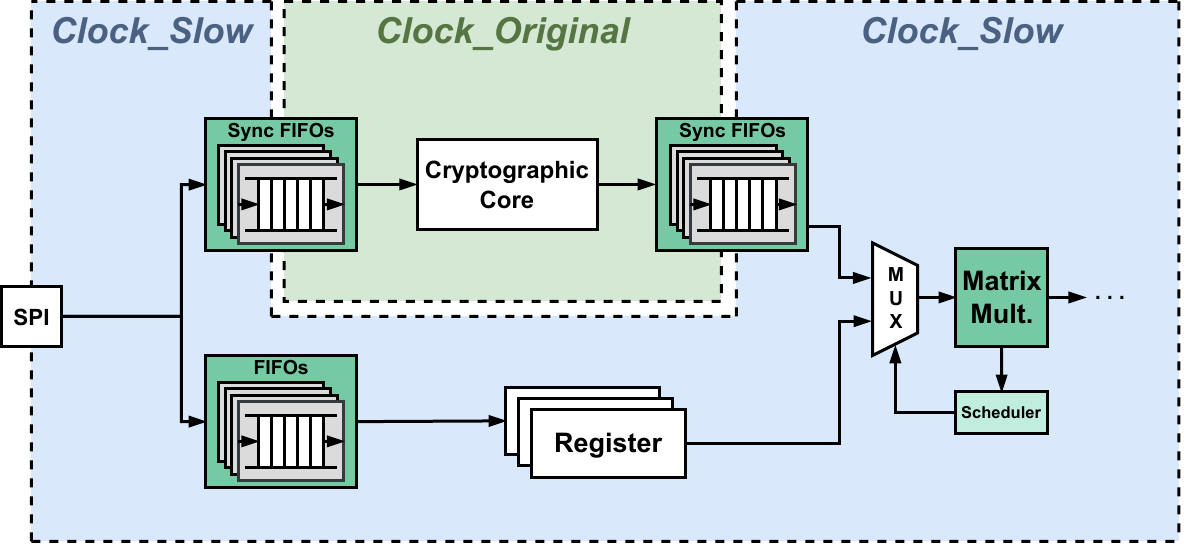}
    \caption{Multi-clock domain architecture}
    \label{fig:multi_clock}
\end{figure}

In addition to serialized matrix multiplication, the system can benefit from other hardware techniques that further refine energy usage and latency, taking full advantage of the current architecture. Given that the ECC core decryption process continues to be the latency bottleneck, even with serialized multiplication, the majority of the system's blocks remain in a static state for extended periods. This static state, characterized by the absence of dynamic switching activity, offers an opportunity for power conservation.

One such technique is clock-gating, which conserves power by deactivating circuits when they are not in use. Another approach involves the implementation of a multi-clock domain system, which assigns different clock speeds to the cryptographic core and the remaining blocks. As depicted in Figure~\ref{fig:multi_clock}, a slower clock($Clock\_Slow$) is generated from the original clock($Clock\_Original$) source, which is then fed to the cryptographic core. The input and output of the cryptographic core require FIFO synchronizers, which use FIFOs to synchronize data being sent across clock domains. All blocks other than the cryptographic core take the original clock source.

While a high-speed clock reduces latency, it also increases power consumption. However, since not all blocks need to operate at the same speed as the ECC core, slower clocks can be assigned to other blocks in the system. This strategy leverages the extended latency of the ECC, allowing other blocks to operate at reduced clock speeds for shorter periods, thereby lowering their power consumption while maintaining the overall system latency.

In this study, we experimented with clocks operating at half and a quarter of the speed of the ECC core clock. However, the multi-clock domain technique did not significantly reduce power and energy consumption, as the ECC operations account for the majority of power usage. Further exploration of this technique could prove beneficial in scenarios where power consumption is more evenly distributed between the cryptographic core and other blocks, potentially yielding more substantial savings.

%\begin{itemize}
%    \item AES-128
%    \item AES-256
%    \item ECC
%    - division of message (pipelining and higher throughput), lower the 
%    \item Possible discussion on Key exchange scenario (not sure if here or later on in paper)
%    \item Lattice-based
%\end{itemize}

\section{Evaluation and Performance Results} \label{section:evaluation and results}

\begin{figure}[b]
    \centering
    \includegraphics[width=1\linewidth]{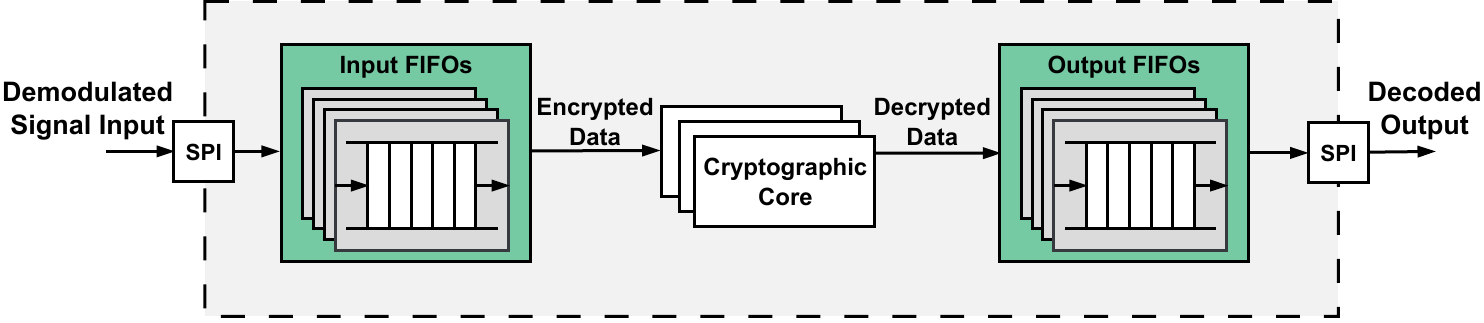}
    \caption{Baseline system architecture}
    \label{fig:system_architecture_baseline}
\end{figure}

\begin{table*}
\centering
\caption{Performance comparison in different channel size}
\label{tab:AES}
\resizebox{\textwidth}{!}{%
\renewcommand{\arraystretch}{2}
\begin{tabular}{|c|c|c|ccc|ccc|c|c|}
\hline
\multirow{2}{*}{} & \multirow{2}{*}{\bf \# of Ch} & \multirow{2}{*}{\bf Block} & \multicolumn{3}{c|}{\bf Area (kGE)} & \multicolumn{3}{c|}{\bf Power (mW)} & {\bf Throughput} & {\bf Energy/bit} \\ \cline{4-9}

 &  &  & \multicolumn{1}{c|}{\bf AES} & \multicolumn{1}{c|}{\bf Mult.} & \bf Inter. Reg. & \multicolumn{1}{c|}{\bf AES} & \multicolumn{1}{c|}{\bf Mult.} & \bf Inter. Reg.  & \bf (Gbps) & \bf (pj/bit)  \\ \hline\hline
 
    % \rowcolor{lightgray}
\bf Baseline & $n$ & $n$-AES & \multicolumn{1}{c|}{17.7} & \multicolumn{1}{c|}{N/A} & N/A & \multicolumn{1}{c|}{1.3*$n$} & \multicolumn{1}{c|}{N/A} & N/A & 0.752*$n$ & 1.72 \\ \hline
\multirow{10}{*}{\bf HUNCC} & 2 & 1-AES 2-Ch. Unmixer & \multicolumn{1}{c|}{\multirow{10}{*}{17.7}} & \multicolumn{1}{c|}{14.1} & 4.3 & \multicolumn{1}{c|}{1.21} & \multicolumn{1}{c|}{0.14} & 0.08 & 1.51 & 0.95 \\ \cline{2-3} \cline{5-11} 
 & 3 & 1-AES 3-Ch. Unmixer & \multicolumn{1}{c|}{} & \multicolumn{1}{c|}{21.4} & 6.4 & \multicolumn{1}{c|}{1.21} & \multicolumn{1}{c|}{0.39} & 0.17 & 2.26 & 0.79 \\ \cline{2-3} \cline{5-11} 
 & 4 & 1-AES 4-Ch. Unmixer & \multicolumn{1}{c|}{} & \multicolumn{1}{c|}{28.2} & 8.7 & \multicolumn{1}{c|}{1.26} & \multicolumn{1}{c|}{0.62} & 0.29 & 3.01 & 0.73 \\ \cline{2-3} \cline{5-11} 
 & 5 & 1-AES 5-Ch. Unmixer & \multicolumn{1}{c|}{} & \multicolumn{1}{c|}{35.6} & 11.6 & \multicolumn{1}{c|}{1.26} & \multicolumn{1}{c|}{0.93} & 0.43 & 3.76 & 0.70 \\ \cline{2-3} \cline{5-11} 
 & 6 & 1-AES 6-Ch. Unmixer & \multicolumn{1}{c|}{} & \multicolumn{1}{c|}{42.7} & 14.6 & \multicolumn{1}{c|}{1.26} & \multicolumn{1}{c|}{1.38} & 0.61 & 4.52 & 0.73 \\ \cline{2-3} \cline{5-11} 
 & 7 & 1-AES 7-Ch. Unmixer & \multicolumn{1}{c|}{} & \multicolumn{1}{c|}{49.8} & 17.6 & \multicolumn{1}{c|}{1.27} & \multicolumn{1}{c|}{1.98} & 0.81 & 5.27 & 0.78 \\ \cline{2-3} \cline{5-11} 
 & 8 & 1-AES 8-Ch. Unmixer & \multicolumn{1}{c|}{} & \multicolumn{1}{c|}{56.5} & 20.0 & \multicolumn{1}{c|}{1.28} & \multicolumn{1}{c|}{2.69} & 1.04 & 6.02 & 0.85 \\ \cline{2-3} \cline{5-11} 
 & 9 & 1-AES 9-Ch. Unmixer & \multicolumn{1}{c|}{} & \multicolumn{1}{c|}{63.6} & 23.9 & \multicolumn{1}{c|}{ 1.19} & \multicolumn{1}{c|}{3.89} & 1.37 & 6.40 & 1.02 \\ \cline{2-3} \cline{5-11} 
 & 10 & 1-AES 10-Ch. Unmixer & \multicolumn{1}{c|}{} & \multicolumn{1}{c|}{69.9} & 26.9 & \multicolumn{1}{c|} {1.09} & \multicolumn{1}{c|}{4.38} & 1.05 & 6.40 & 1.1 \\ \cline{2-3} \cline{5-11} 
 & 11 & 1-AES 11-Ch. Unmixer & \multicolumn{1}{c|}{} & \multicolumn{1}{c|}{76.5} & 30.9 & \multicolumn{1}{c|} { 0.98} & \multicolumn{1}{c|}{5.18} & 1.74 & 6.40 & 1.25 \\ \hline
\end{tabular}%
}
\end{table*}

This section showcases the performance of the energy-efficient HUNCC, equipped with different cryptographic cores, as developed in this study. The system was designed and implemented using 28nm CMOS technology, considering various scenarios, corresponding hardware architectures, and techniques. The power, area, and throughput results were derived from post-synthesis simulation operating at 0.9V and 100MHz clock frequency.

Figure~\ref{fig:system_architecture_baseline} presents the baseline system used for comparison with our proposed system. This baseline system, similar to our system, supports the same number of channels, with each channel furnished with its own cryptographic core to ensure robust computational security. However, unlike our system, the baseline system lacks a multiplication block and intermediate registers, focusing solely on the cryptographic process. Both the baseline and our proposed system were synthesized and subjected to post-synthesis simulation under the same conditions. This configuration provides a comparative baseline to highlight the efficiency, performance enhancements, and throughput improvements introduced by our proposed system.

\subsection{ HUNCC with AES Cryptographic Core} \label{results:AES}

\begin{figure*}[!t]
    \centering
    \begin{subfigure}[b]{0.475\textwidth}
        \centering
        \includegraphics[width=\textwidth]{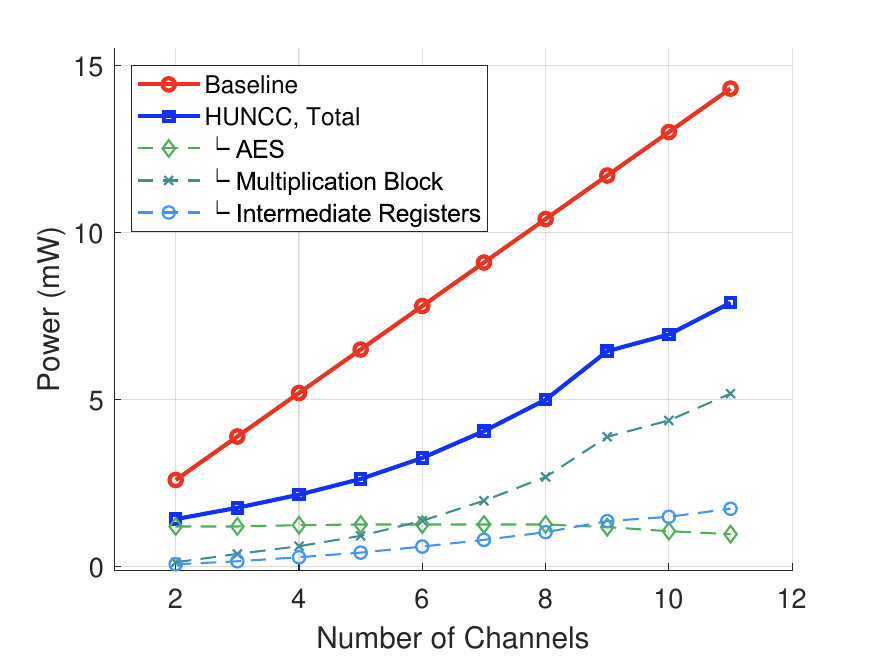}
        \caption[Power - Redraw with highlighting HUNCC and AES and then with lower preference for breakdowns]%
        {{\small Power}}    
        \label{fig:power_breakdown}
    \end{subfigure}
    \hfill
    \begin{subfigure}[b]{0.475\textwidth}  
        \centering 
        \includegraphics[width=\textwidth]{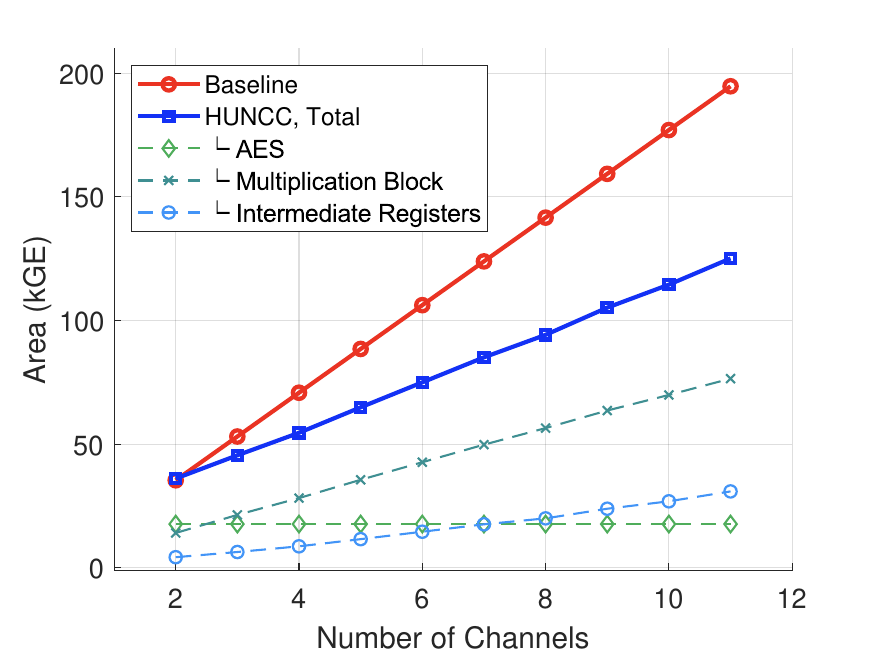}
        \caption[]%
        {{\small Area}}    
        \label{fig:area_breakdown}
    \end{subfigure}
    \vskip\baselineskip
    \begin{subfigure}[b]{0.475\textwidth}   
        \centering 
        \includegraphics[width=\textwidth]{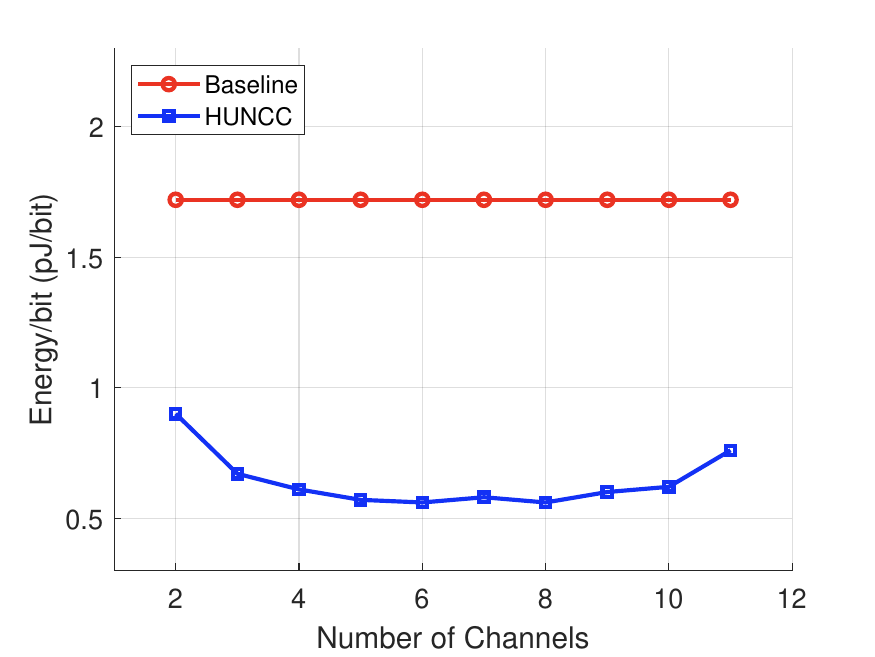}
        \caption[]%
        {{\small Energy/bit}}    
        \label{fig:energy_per_bit_comparison}
    \end{subfigure}
    \hfill
    \begin{subfigure}[b]{0.475\textwidth}  
        \centering 
        \includegraphics[width=\textwidth]{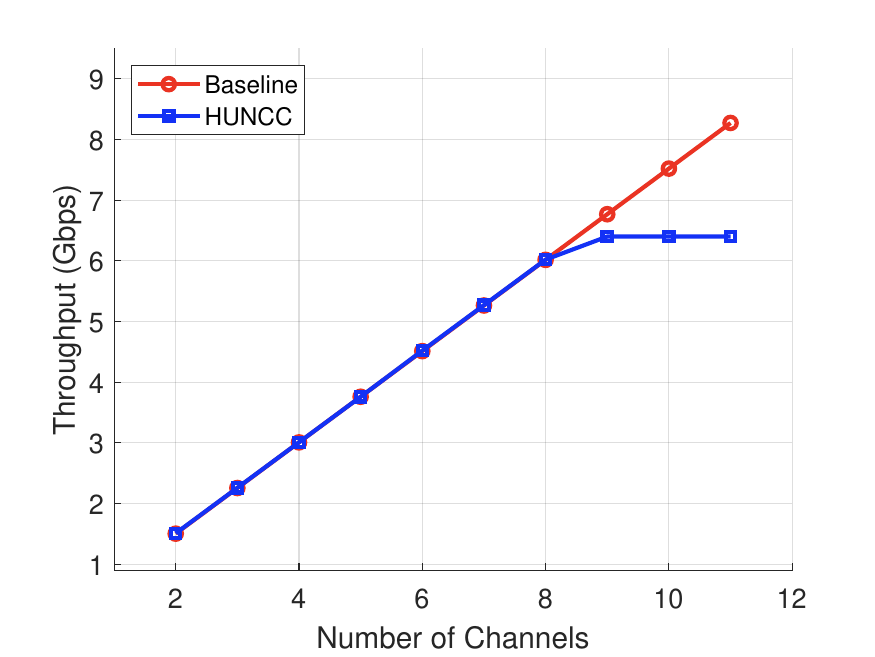}
        \caption[]%
        {{\small Throughput}}    
        \label{fig:throughput_comparison}
    \end{subfigure}
    \caption[ The average and standard deviation of critical parameters ]
    {\small Power, throughput, energy/bit and area comparison between baseline and proposed system} 
    \label{fig:AES_HUNCC_comparison}
\end{figure*}

In our study, we simulated the proposed HUNCC with AES-256 and GF($2^{16}$) across various channel configurations. Table~\ref{tab:AES} provides a comparative analysis of power, throughput, decoding energy/bit, and area between the baseline system, and our proposed system. The baseline system, designed with an AES core for each channel, exhibits a linear increase in power, throughput, and area. Notably, the decoding energy/bit remains consistent across all channel configurations in the baseline system.

In contrast, our proposed system incorporates a single AES core, complemented by a multiplication block and intermediate registers whose sizes vary depending on the channel count. This design results in a slower power increase compared to the baseline system, while maintaining equivalent throughput up to 8-channel scenarios. Consequently, the decoding energy/bit in our system is significantly lower. For instance, at 5 channels, our system achieves a decoding energy/bit of 0.7 pj/bit, compared to the baseline's 1.72 pj/bit.

Despite the integration of an additional multiplication block and intermediate registers post-AES decryption, our system maintains throughput parity with the baseline. This is achieved by pipelining all processes, allowing multiplication to be completed within the AES decryption process. Both systems exhibit a linear increase in area as the channel count grows, but the increase is more pronounced in the baseline system due to the complexity of the AES core architecture.

\begin{table}[]

%\resizebox{\columnwidth}{!}{%
\centering    
\resizebox{\linewidth}{!}{%
\begin{threeparttable}
\centering    
    \caption{\centering HUNCC vs. state-of-the-art AES}
    \label{tab:AES_ref}
\renewcommand{\arraystretch}{2}
%\resizebox{\columnwidth}{!}{%
\begin{tabular}{|c|c|c|c|c|c|}
\hline
 &\bf  \cite{1690090} \tnote{a} &\bf  \cite{Mathew201053GbpsNG} & \bf  \cite{7019004} & \bf  \cite{7573553} & \bf  This work \tnote{a} \\ \hline
 
\renewcommand{\arraystretch}{1}
\bf Security Type & \begin{tabular}[c]{@{}c@{}}AES-128\\ Comp.\end{tabular} & \begin{tabular}[c]{@{}c@{}}AES-256\\ Comp.\end{tabular} & \begin{tabular}[c]{@{}c@{}}AES-128\\ Comp.\end{tabular} & \begin{tabular}[c]{@{}c@{}}AES-128\\ Comp.\end{tabular} & \begin{tabular}[c]{@{}c@{}}AES-256\\ Comp. \& Ind. \end{tabular} \\ \hline
\bf Operation Mode & Encrypt & Both & Decrypt & Encrypt & Decrypt \\ \hline

\bf Technology (nm) & 130 & 45 & 22 & 40 & 28 \\ \hline 
\bf Area (kGE) & 3.2 & N/A & 2.1 & 2.2 & 64.9 \\ \hline
\bf Cycles/Block & 160 & 5 & 216 & 336 & 17 \\ \hline
\bf Throughput \tnote{b} (Gbps) & 0.078 & 1.8 & 59.22 & 0.038 & 3.76 \\ \hline
\bf Voltage (V) & 1.2 & 1.1 & 0.9 & 0.9 & 0.9 \\ \hline
\bf Energy/bit (pJ/bit) & 37.5 & 2.3 & 19.5 & 8.9 & 0.7 \\ \hline
\end{tabular}%}

     \begin{tablenotes} \footnotesize
       \item [a] Post-synthesis area and power
       \item [b] Normalized to 100MHz clock frequency
     \end{tablenotes}
  \end{threeparttable}
%  }
}

\end{table}

Figure~\ref{fig:AES_HUNCC_comparison} provides a detailed breakdown of the power and area contributions from the AES core, multiplication block, and intermediate registers within the system, across a range of channel counts. At the outset, when the channel count is at 2, the AES core dominates both power and area. However, as the number of channels increases, the multiplication block and intermediate registers progressively take on a more significant role, becoming the dominant factor in both power and area, while the AES core's contributions remain steady.

%Interestingly, a shift in throughput dynamics is observed when the channel count exceeds 8. At this point, the latency bottleneck transitions from the AES core to the multiplication block. This shift results in the AES core experiencing periods of idle time, as it must wait for the slower multiplication block to complete its operations. Consequently, this idle time reduces the dynamic power consumption of the AES core, as it is no longer in constant operation. This latency bottleneck transition to the multiplication block limits the throughput of our HUNCC, and the limited throughput leads to an increase in energy/bit.

A notable shift in throughput dynamics occurs when the channel count surpasses 8, as depicted in Figure~\ref{fig:throughput_comparison}. Up to this point, both the baseline and HUNCC maintain equivalent throughput, a testament to the efficiency of our pipelining and parallelization strategies. However, as the channel count exceeds 8, the throughput of HUNCC becomes constrained, while the baseline's throughput continues to ascend. This shift stems from the transition of the latency bottleneck from the AES core to the multiplication block within HUNCC. Consequently, the AES core experiences periods of idle time, waiting for the slower multiplication block to complete its operations. This idle time leads to a reduction in the dynamic power consumption of the AES core, as it is no longer continuously operational. This reduction is depicted in Figure~\ref{fig:power_breakdown}. As a result, the transition of the latency bottleneck to the multiplication block restricts the throughput of HUNCC, leading to an increase in energy/bit.

% \begin{figure}%[!t]
%    \centering
%    \includegraphics[width=1\linewidth]{figures/comparison_with_ref.jpg}
%    \caption{Comparison of our HUNCC with AES with state of the art}
%    \label{fig:comparison_with_ref}
% \end{figure}

% \begin{figure}%[!t]
%    \centering
%    \includegraphics[width=1\linewidth]{figures/comparison_with_ref_throughput_normalized.jpg}
%    \caption{Comparison of our HUNCC with AES with state of the art}
%    \label{fig:comparison_with_ref_throughput_normalized}
% \end{figure}

% \usepackage{graphicx}

% \footnotesize{$^a$ The smallest spatial unit is county, $^b$ more details in appendix A}\\

%Table~\ref{tab:AES_ref} presents a comparison between our HUNCC equipped with AES-256 and other state-of-the-art AES accelerators. Our work offers the same level of computational security as AES-256, but also provides individual security inherent to the HUNCC design. Despite the larger area dominated by the unmixing block, our architecture achieves lower energy consumption per bit and delivers a throughput in the order of Gbps. The flexible nature of our system allows for the integration of AES accelerators from \cite{1690090,Mathew201053GbpsNG,7019004,7573553}, potentially enhancing the energy efficiency per bit.

Table~\ref{tab:AES_ref} presents a comparison between our HUNCC equipped with AES-256 and other state-of-the-art AES accelerators. Our system offers the same level of computational security as AES-256, while also providing the individual security inherent to the HUNCC design. Despite the larger area dominated by the unmixing block, our architecture achieves lower energy consumption per bit and delivers a throughput in the order of Gbps.

It's important to acknowledge that the differences in technology, voltage, and the nature of the results (fabricated chip versus post-synthesis simulation) can influence the exact figures for area and energy/bit \cite{782564}. However, these variations do not diminish the value of our work. The inherent flexibility of HUNCC allows for the integration of a wide range of cryptographic protocols, potentially leading to significant energy savings per bit while maintaining the same level of security. The potential for integrating AES accelerators from \cite{1690090,Mathew201053GbpsNG,7019004,7573553} further demonstrates the versatility and energy efficiency of our HUNCC design.

In conclusion, while the technology differences, voltage variations, and the nature of the results may impact the specific metrics, the overall trends and comparisons provide valuable insights. The adaptability of our design, as demonstrated by the potential integration of various AES accelerators, underscores the robustness and flexibility of HUNCC.

\subsection{ HUNCC with Elliptic Curve Cryptographic Core} \label{results:ECC}

\begin{table*}[!bh]
\centering
    \caption{\centering Performance comparison in 16-Ch ECC system}
    \label{tab:ECC}
\renewcommand{\arraystretch}{2}
%\resizebox{\width\textwidth}{!}{%
\begin{tabular}{|c|c|cccc|cccc|}
\hline
 &
  \textbf{16-Ch ECC} &
  \multicolumn{4}{c|}{\textbf{16-Ch HUNCC with ECC}} &
  \multicolumn{4}{c|}{\textbf{\begin{tabular}[c]{@{}c@{}}16-Ch HUNCC with ECC\\ Serialized Multiplication\end{tabular}}} \\ \hline
\textbf{Block} &
  16-ECC &
  \multicolumn{1}{c|}{1-ECC} &
  \multicolumn{1}{c|}{Mult.} &
  \multicolumn{1}{c|}{Inter. Reg.} &
  \textbf{Total} &
  \multicolumn{1}{c|}{1-ECC} &
  \multicolumn{1}{c|}{Mult.} &
  \multicolumn{1}{c|}{Inter. Reg.} &
  \textbf{Total} \\ \hline
\textbf{Power (mW)} &
  42.72 &
  \multicolumn{1}{c|}{2.64} &
  \multicolumn{1}{c|}{0.032} &
  \multicolumn{1}{c|}{0.019} &
  \textbf{2.691} &
  \multicolumn{1}{c|}{2.745} &
  \multicolumn{1}{c|}{0.002} &
  \multicolumn{1}{c|}{0.069} &
  \textbf{2.816} \\ \hline
\textbf{Area (kGE)} &
  1707 &
  \multicolumn{1}{c|}{102} &
  \multicolumn{1}{c|}{222} &
  \multicolumn{1}{c|}{80} &
  \textbf{404} &
  \multicolumn{1}{c|}{102} &
  \multicolumn{1}{c|}{0.9} &
  \multicolumn{1}{c|}{80} &
  \textbf{182.9} \\ \hline
\textbf{Throughput (Kbps)} &
  4934 &
  \multicolumn{4}{c|}{4934} &
  \multicolumn{4}{c|}{4934} \\ \hline
\textbf{Energy/bit (pJ/bit)} &
  8656 &
  \multicolumn{4}{c|}{564} &
  \multicolumn{4}{c|}{593} \\ \hline
\end{tabular}%
%}
\end{table*}

Our proposed system's adaptability enables the integration of an ECC core. Table~\ref{tab:ECC} offers a comparative analysis of power, area, throughput, and energy/bit among three distinct systems: a baseline system solely equipped with ECC cores, HUNCC incorporating ECC, and HUNCC integrating ECC with an additional serialized multiplication feature. All these systems operate with 16 channels in GF($2^{16}$), the maximum limit imposed by the field size.

ECC cryptography, being more power-demanding and exhibiting higher latency compared to hardware multiplication, results in the multiplication process consuming a relatively minor portion of the total power. This characteristic allows us significant power savings while maintaining consistent throughput, leading to a 14.6× reduction in decoding energy/bit when compared to the baseline system of 16 independent ECC cores. This substantial energy saving potentially positions ECC as a more viable option for encryption and decryption processes, which were previously less favored due to their high energy and latency costs. Consequently, ECC can now provide individual security in a more energy-efficient manner.
In the context of serialized multiplication techniques, there is a marginal increase in power consumption in multiplication block and intermediate registers due to elevated internal and switching power. However, this increase is negligible when compared to the power consumption of the ECC core. The most notable benefit emerges in the form of area reduction. As the number of channels escalates, the system area becomes increasingly dominated by the multiplication block. By implementing serialized multiplication, we can significantly shrink this block's size, resulting in a 9.3× saving in area compared to the baseline system of 16 independent ECC cores.

In conclusion, the proposed system underscores the potential for substantial improvements in energy efficiency and area conservation when employing ECC in HUNCC. The system design's flexibility facilitates the integration of diverse cryptographic cores and the application of various hardware techniques, thereby emphasizing the proposed system's adaptability and potential.

\section{Discussions} \label{discussions}
%Post-quantum applications, having one secure link is good enough for providing individual security for all links – MI leakage – compression to achieve enough randomness - Side channel attacks, key based attacks, brute forcing

The hardware architecture we propose has numerous advantages over a traditional cryptographic core design. The universality and flexibility of this scheme make it an option for energy reduction for any cryptosystem when one has more than a single unit of message block to be encrypted. This makes a secure encryption scheme possible for resource-constrained IoT devices. The higher the number of messages that are mixed in the precoding stage, the better the energy efficiency of the proposed scheme. There are a few limiting factors in mixing a large number of messages. One of the limitations comes from the field size of operations. As discussed in \cite{silva2009universal}, the size of the MRD code that can securely transmit messages when at least one channel is protected from the Eve is limited by the field size of the coding operations. Most cryptosystems also works on specific Galois fields, but the field size of precoding the coding operations need not be matched to that of the crypto operations, but it can not be increased unbounded due to the complexity and implementation costs that are proportional to the field size. However, as we saw in the previous section, a field size of $2^{16}$ is practically feasible and provides a significant reduction in energy per bit (2.5× in AES and 14.6× in ECC). It is also notable that in an optimized cryptosystem with low operational cost, the number of input channels to be mixed is sometimes limited by other factors such as throughput or energy per bit when the unmixer operations take more clock cycles than the decryption or get costlier than the decryption respectively. Such trade-offs are clearly visible in the AES-256 discussion. 

The architecture around ECC shows that when an expensive cryptosystem is to be used as the core, the HUNCC architecture provides more than one avenue of optimization. This shows the flexibility of the system and its suitability to operate with expensive post-quantum cryptosystems. The public key cryptosystems are used primarily for key exchange than sending information due to its high energy requirement. It is noteworthy that the significant reduction in the energy per bit for the expensive public key cryptosystem also makes it more suitable to use it directly for sending information. This makes it more secure compared to a symmetric key cryptosystem without increasing power consumption. Even though our hardware devices were focused on the receiver side, the coding operations on the sender side are also very similar, as shown in Section \ref{sec:system}. For a traditional public-key cryptosystem where the encryption operation is more costly, the proposed approach can provide a significant reduction in energy cost. However, IoT devices are already employing lightweight cryptosystems based on ECC. Given that ECC operations are fundamentally based on elliptic curve scalar multiplication (ECSM), both encryption and decryption operations bear a similar cost \cite{singh2015implementation}. The encryption operation in symmetric cryptosystems such as the AES-256 is also very similar to the decryption in their corresponding cryptosystems, which can achieve a similar improvement in the performance as we see on the receiver side.

The security analysis of HUNCC has been presented in detail in \cite{cohen2021network,cohen2022partial}. It is proven that the proposed system provides individual secrecy for the messages when an eavesdropper is not observing all the channels (in a multi-channel communication scenario) while providing the same computational security level as the underlying cryptosystem when at least one channel is encrypted. Further, it is also proven that if the cryptosystem used is IND-CCA1 secure, the energy efficient cryptosystem is Individually IND-CCA1 secure. It is also to be noted that the precoding matrix used in the HUNNC system can be made available to the public without having any impact on security. The unmixer component of HUNCC cannot provide any meaningful information about individual messages unless the decryption is executed correctly. This unique characteristic of HUNCC provides a significant advantage in terms of security. For example, since it doesn't rely on additional keys beyond AES, it is not vulnerable to key-based attacks \cite{isa2011aes}. Consequently, HUNCC is as secure as AES against key-based attacks. The linear mixing of packets before encoding may however leak patterns of the input distribution. It is expected that the input distribution is uniform to assure security. 

\section{Conclusions} \label{conclusions}
%A comprehensive discussion on energy efficient cryptosystems - Around 1pJ/bit AES - Above 90 \% reduction with ECC - Makes costly cryptosystems to be used for practical purposes 

%\textcolor{red}{[connect to the IoT storyline, expand with the summary]} 
This paper presents a comprehensive study of the hardware architecture for an energy-efficient cryptosystem, CERMET,, along with its optimizations. The novel approach of coding over multiple channels and encrypting only one of them achieves the same level of computational security as the underlying cryptosystem. This innovation makes otherwise prohibitively energy-hungry cryptosystems practical without incurring significant operational costs. Furthermore, it reduces the energy requirements of decryption to less than 1 pJ/bit, making it comparable to modern error control decoding. The universality of the approach is demonstrated by presenting both symmetric and asymmetric cryptosystems. Various hardware optimization techniques, such as efficient Galois field multiplications, pipelining to reduce idle time, and serialization or parallelization of multiplications depending on the crypto operations, are also explored in detail as part of the study. CERMET provides a highly modularized architectural framework where different combinations of these optimizations can be used depending on the application. 

The optimal design for an AES-256 implementation using CERMET in 28nm technology achieves less than 1 pJ/bit energy efficiency, marking a 2.5× energy per bit reduction compared to the baseline system with the identical AES-256 accelerator. This energy efficiency surpasses even the most efficient error control decoding schemes in the literature. When analyzing a more computationally intensive cryptographic protocol, such as a public key cryptosystem based on ECC, it is shown that a substantial 14.6× reduction in energy per bit and a 9.3× reduction in area can be achieved. This significant improvement makes such cryptosystems much more suitable for low-energy use cases.


\ifCLASSOPTIONcompsoc
  %\section*{Acknowledgments}
\else
  %\section*{Acknowledgment}
\fi


\bibliographystyle{IEEEtran}
\bibliography{references.bib}

\appendices

\renewcommand{\thesubsection}{\Alph{subsection}}
\section{HUNCC Algorithm} \label{appendix:HUNCC}

In section \ref{sec:system}, we presented the hybrid cryptosystem HUNCC which is illustrated in Figure \ref{fig:HUNCC_overview}. Here, for completeness, we give a more detailed explanation of HUNCC algorithm \cite{cohen2021network}, as presented in Algorithm \ref{alg:hybrid_scheme}. We start with the encoding process at Alice. Let $(\mathrm{Enc},\mathrm{Dec},p_k,s_k)$ be a cryptosystem, as described in Definition \ref{Crypto_scheme}, with security level $b$. Alice chooses a number $c$ of input channels to be encrypted. Without loss of generality, we let the input channels indexed by $1,\ldots,c$ be the encrypted ones. Let the total number of channels, $n$, be fixed, and consider blocks of messages $M^{(1)}, \ldots,M^{(N)}$, where each $M^{(i)} = [M_1^{(i)}, \ldots, M_{n}^{(i)}]$ with $M_j^{(i)} \in \mathbb{F}_{2^n}$ is generated independently, and uniformly at random. Let $\textbf{G} \in \mathbb{F}_{2^m}^{n \times n}$ be an MRD secure linear code as in Section \ref{sec:system}. Thus, the vectors $X^{(1)}, \ldots, X^{(N)}$, where $X^{(i)} =\textbf{G} M^{(i)} $ correspond to an MRD secrecy encoding of $M^{(i)}$ (c.f. lines 1-5 in Algorithm~\ref{alg:hybrid_scheme}).

\begin{algorithm} [t]
  \caption{: HUNCC Scheme \cite{cohen2021network}}
  \label{alg:hybrid_scheme}
  \begin{algorithmic}[1]
    \Statex\textbf{Input:}  At Alice, $n$ messages $[M_1;\ldots;M_{n}] \in F^{n}_{q^m}$ %of length $\kb$ bits each over $\mathbb{F}_{2}$

    \Statex\textbf{\underline{Encoding scheme at Alice}:}
    \State \textbf{Stage 1:} Individual MRD secrecy encoding
    \State For the $n$ messages in a block $[M^{(1)}, \ldots, M^{(N)}] \in \mathbb{F}_{2^{m}}$
    \ForEach{column $i\in\{1,\ldots,N\}$ in the block  $[M^{(1)}, \ldots,M^{(N)}]$}
        \State $X^{(i)} =  \textbf{G} M^{(i)}$
    \EndFor
    \State \textbf{Stage 2:} Key encryption
    \ForEach{input channel $1\leq i \leq c$ which support cryptosystem}
        \State $\ddot{\mathbf{b}}_{i}\in \mathbb{F}_2^{k_{in}} \leftarrow [X_{i}^{(1)},\ldots, X_{i}^{(N)}] \in \mathbb{F}_{2^{m}}$
        \State $\mathbf{y}_i = \mathrm{Enc}(\ddot{\mathbf{b}}_i, p_i)$
    \EndFor

    \Statex\textbf{\underline{Transmission over the multipath network}:}
    \ForEach{input channel $1\leq i \leq c$ which support cryptosystem}
        \State Transmit $\mathbf{y}_i$
    \EndFor
    \ForEach{remaining input channels $1\leq i \leq (n-c)$ without cryptosystem}
        \State Transmit $\mathbf{y}_i = [X_{i}^{(1)},\ldots, X_{i}^{(N)}] \in \mathbb{F}_{2^{m}}$
    \EndFor

    \Statex\textbf{\underline{Decoding scheme at Bob}:}
    \State \textbf{Stage 1:} Key decryption
    \ForEach{output channel $1\leq i \leq c$ with cryptosystem}
        \State $[X_{i}^{(1)},\ldots, X_{i}^{(N)}] \leftarrow \mathrm{Dec}(\mathbf{y}_i, s_i)$
    \EndFor
    \State \textbf{Stage 2:} Individual MRD secrecy decoding
    \ForEach{column $i\in\{1,\ldots, N\}$ in the block  $[X^{(1)}, \ldots,X^{(N)}]$}
        \State $(M_{1}^{i};\ldots;M_{n}^{i}) = \textbf{H} X^{i}$
    \EndFor
    \Statex\textbf{Output:} At Bob, $[M_1;\ldots;M_{n}] \in F^{n}_{q^m}$
  \end{algorithmic}
\end{algorithm}

For every input channel $i \in c$, consider the collection of symbols $X_{i}^{(1)},\ldots, X_{i}^{(N)}$. Since $\mathbb{F}_{2^n} \simeq \mathbb{F}_2^{n}$, the collection $X_{i}^{(1)},\ldots, X_{i}^{(N)}$ can be injectively mapped into a sequence of bits $\ddot{\mathbf{b}}_{i}$ of length of an individual input ($k_{in}$). Each $\ddot{\mathbf{b}}_i$ is encoded via the key encryption before being sent, i.e., each input channel $i$ transmits $\mathbf{y}_i = \mathrm{Enc}(\ddot{\mathbf{b}}_i, p_i)$. Note that $\mathbf{y}_i$ is of length $n$ (c.f. lines 6-13 in Algorithm~\ref{alg:hybrid_scheme}). For the input channels $i > c$, Alice directly sends the collection $X_{i}^{(1)},\ldots, X_{i}^{(N)}$ unencrypted (c.f. lines 14-16 in Algorithm~\ref{alg:hybrid_scheme}).

Now, we detail the decoding process at Bob. For the $c$ encrypted paths, Bob uses the private key to decode the messages (c.f. lines 17-20 in Algorithm~\ref{alg:hybrid_scheme}). Thus, Bob obtains $[X_{i}^{(1)},\ldots, X_{i}^{(N)}] = \mathrm{Dec}(\mathbf{y}_i, s_i)$ for every $i \in c$. The messages obtained via the remaining $n-c$ paths were unencrypted. Thus, together with the decrypted messages from the first $c$ paths, Bob has the entirety of $X^{(1)}, \ldots,X^{(N)}$. Now, for each $i$-th column in the estimated encoded block, Bob uses the parity check $\textbf{H} \in \mathbb{F}_{2^m}^{n \times n}$ to obtain the original messages transmitted (c.f. lines 21-24 in Algorithm~\ref{alg:hybrid_scheme}).

\section{Multiplication Algorithm}
\label{appendix:mult_algorithm}

In this appendix, we provide a detailed explanation of the Russian Peasant Algorithm (RPA) for Galois field Multiplication \cite{gimmestad_1991}, as presented in Algorithm \ref{alg:RPA}. The RPA is a fast multiplication algorithm that operates on two input polynomials $a$ and $b$, along with an irreducible polynomial $p$ of degree $m$. The algorithm begins by checking if either of the input polynomials is zero, in which case the result is zero. It then proceeds with a bitwise operation, shifting and exclusive OR-ing the polynomials based on the least significant bit of $b$. If the high bit of $a$ is set, $a$ is XOR-ed with $p$. This process is repeated for $m$ iterations, resulting in the product $c$ of the two input polynomials under the Galois field defined by $p$.

\begin{algorithm}[tb]
\caption{: Russian Peasant Algorithm for Galois Field Multiplication~\cite{gimmestad_1991}}
\label{alg:RPA}
\begin{algorithmic}[1]
\Statex \textbf{Input:} 
\Statex $a$ : the first polynomial
\Statex $b$ : the second polynomial
\Statex $p$ : the irreducible polynomial
\Statex $m$ : the degree of the irreducible polynomial
\If{$a = 0$ or $b = 0$}
    \State \Return 0
\EndIf
\State mask $\gets 1 << (m-1)$
\State highbit $\gets 0$
\State c $\gets 0$
\For{$i \gets 0$ to $m-1$}
    \If{$b = 0$}
        \State \textbf{break}
    \EndIf
    \If{$b$ AND $1$}
        \State c $\gets c \oplus a$
    \EndIf
    \State highbit $\gets a$ AND mask
    \State a $\gets a << 1$
    \State b $\gets b >> 1$
    \If{highbit}
        \State a $\gets a \oplus p$
    \EndIf
\EndFor
\State \Return c
\end{algorithmic}
\end{algorithm}

\begin{algorithm}
\caption{: Galois Field Multiplication using Log and Exponential Look-up Tables \cite{10.1007/3-540-47555-9_14}}
\label{alg:log_exp}
\begin{algorithmic}[1]
\Statex \textbf{Input:}
\Statex $a$ : the first polynomial
\Statex $b$ : the second polynomial
\Statex $p$ : the primitive polynomial
\Statex $m$ : the degree of the primitive polynomial
\Statex\textbf{\underline{Log and Exponential Table Creation}:}
    \State order $\gets 1 << m$
    \State log, exponential $\gets$ [None]*order, [None]*order
    \State power $\gets 1$
\For{$i \gets 0$ to $order-2$}
    \State log[power] $\gets i$
    \State exponential[i] $\gets power$
    \State power $\gets$  RPA\_multiply(2, power, p, m)
\EndFor

\Statex\textbf{\underline{Multiplication using Log and Exponential tables}:}
\If{$a = 0$ or $b = 0$}
    \State \Return 0
\EndIf
\State exp\_sum $\gets (log[a] + log[b]) \% (1 << m - 1)$
\State \Return exponential[exp\_sum]
\end{algorithmic}
\end{algorithm}

Next, we describe the Galois field multiplication using log and exponential look-up tables \cite{10.1007/3-540-47555-9_14}, as presented in Algorithm \ref{alg:log_exp}. This algorithm also operates on two input polynomials $a$ and $b$, along with a primitive polynomial $p$ of degree $m$. The algorithm first constructs the log and exponential look-up tables based on the primitive polynomial $p$. The log table records the exponent corresponding to each power of 2, while the exponential table records the power of 2 corresponding to each exponent. These tables are constructed using the RPA for multiplication. Once the tables are constructed, the multiplication of $a$ and $b$ is performed by adding their corresponding log values, taking the modulus with respect to the order of the field, and then looking up the result in the exponential table. This approach significantly speeds up the multiplication process in the Galois field defined by $p$.

These two algorithms, RPA and Log/Exponential multiplication, provide efficient methods for performing multiplication in Galois fields, which are crucial for many cryptographic and error correction codes. They are not directly derived from the reference papers but are inspired by the concepts and techniques presented therein. The algorithms have been adapted and optimized for the specific requirements of our system.



%\bibliographystyle{IEEEtran}

%\begin{thebibliography}{1}
%\bibitem{lidl_niederreiter_1996}
%Lidl, R., & Niederreiter, H. (1996). Finite Fields (2nd ed., Encyclopedia of Mathematics and its Applications). Cambridge: Cambridge University Press. doi:10.1017/CBO9780511525926
%\end{thebibliography}

% \bibliographystyle{IEEEtran}
% \bibliography{references.bib}

\end{document}